\documentclass[12pt]{iopart}

\usepackage{graphicx}

\begin{document}

\title[Painting arbitrary and dynamic potentials for BECs]{Experimental demonstration of painting arbitrary and dynamic potentials for Bose-Einstein condensates}

\author{K Henderson, C Ryu, C MacCormick\footnote{Present address: Department of Physics and Astronomy, The Open University, Walton Hall, Milton Keynes, MK7 6AA, UK; c.maccormick@open.ac.uk.} and M G Boshier}

\address{Physics Division, Los Alamos National Laboratory, Los Alamos, NM 87545, USA}

\ead{boshier@lanl.gov}

\begin{abstract}
There is a pressing need for robust and straightforward methods to create potentials for trapping Bose-Einstein condensates which are simultaneously dynamic, fully arbitrary, and sufficiently stable to not heat the ultracold gas.  We show here how to accomplish these goals, using a rapidly-moving laser beam that ``paints'' a time-averaged optical dipole potential in which we create BECs in a variety of geometries, including toroids, ring lattices, and square lattices.  Matter wave interference patterns confirm that the trapped gas is a condensate.  As a simple illustration of dynamics, we show that the technique can transform a toroidal condensate into a ring lattice and back into a toroid. The technique is general and should work with any sufficiently polarizable low-energy particles.
\end{abstract}


\section{Introduction}
An atomic Bose Einstein condensate (BEC), which can be thought of as a large number of atoms mostly occupying the same single particle state, is one of the most fundamental quantum many-body systems.  This essential simplicity is responsible for a huge body of research on atomic BEC, including topics such as matter-wave interferometry, quantum information processing, superfluidity, many-body physics, and quantum phase transitions.  Most of the experimental work in these areas introduces dynamics to the system through changes in the potential trapping the atoms, which is either an inhomogeneous magnetic field (coupling to the atomic magnetic dipole moment) or an inhomogeneous off-resonant laser field (coupling to the atomic polarizability), or a combination of the two.   Much effort has naturally gone into attempts to develop potentials which are simultaneously dynamic, fully arbitrary, and sufficiently smooth and stable to not heat the ultracold BEC.  However, there are still no robust and easily-implemented methods available for producing such potentials.  We show here how to accomplish these goals, using a rapidly-moving laser beam that ``paints'' a time-averaged optical dipole potential in which we create and manipulate BECs in a variety of geometries.

This work is partly motivated by several theoretical proposals concerning the physics of quantum degenerate gases which will now be accessible to experiment.  One example is the theoretical attention currently being directed at quantum gases in multiply-connected geometries such as toroids and ring lattices.  These geometries deliver freedom from end effects, they realize periodic boundary conditions, and they can stabilize topological defects such as vortices.  Even relatively straightforward manipulations of a simple toroidal BEC would implement several proposals \cite{Brand2001, Nugent2003} going beyond the recent first demonstration of persistent currents in a BEC confined in a magnetic trap with an optical barrier at the centre \cite{Ryu}.  The toroidal geometry, where winding number is a robustly conserved quantity, is also ideal for studying topological defects produced in a rapid quench through the BEC phase transition \cite{Anglin1999, Weiler2008}.  There are intriguing proposals for ring lattices \cite{Amico2005} and, very recently, for exploring the non-equilibrium dynamics of quantum phase transitions in the same geometry \cite{Dziarmaga2008}.  The ring lattice condensate may also offer a path to creation of macroscopic quantum superposition states \cite{Rey2007}.   Finally, a method to create arbitrary potential lattices is necessary for quantum simulation and studies of so-called tailored matter \cite{Lewenstein2007} and it is desirable for quantum information processing \cite{Beugnon2007}.

\section{Background}
Magnetic trapping technology was used to create potentials in the pioneering BEC experiments, either as the time-averaged, orbiting potential (TOP) trap \cite{Anderson1995} or as the static Ioffe-Pritchard trap \cite{Mewes1996}. Although the relatively large distance from the BEC to the current-carrying conductors in these systems guarantees smooth potential surfaces, the fixed conductors also limit the possible field configurations.  Some flexibility in potential geometry can be gained by superimposing optical dipole potentials to create hybrid traps \cite{Davis1995} or by imposing additional magnetic fields \cite{Arnold2002}.  The possibilities of magnetic trapping expanded further with the advent of surface microtraps using more complicated and multiplexed electrodes \cite{Fortagh}, and they were extended again by the addition of RF-dressing \cite{Zobay, Schumm2005}.   Recently an RF-dressed magnetic trap was combined with fixed light sheets to realize a condensate ring trap \cite{Heathcote2008}.  However, constrained potential geometry remains an unavoidable limitation of magnetic trapping with fixed conductors.

Optical dipole potentials are considerably more flexible because lasers can be tightly focused and easily directed at a BEC from outside of the high vacuum environment containing the condensate.  Most work in this area has used acousto-optic deflectors (AODs) to provide rapid deflection and modulation.  Rapidly-moving blue-detuned laser beams have realized mirrors for BECs \cite{Bongs1999}, and traps \cite{Friedman2000} and potential barriers \cite{Milner2001, Friedman2001} for cold atoms.  A BEC was transferred to three or four multiplexed optical traps in a static configuration \cite{Onofrio}, and the same group later used multiple-frequency driving of an AOD \cite{Shin} to realize the first BEC interferometer.  More recently, spatial light modulator (SLM) technology (widely employed in biological and microparticle optical tweezers, see e.g. \cite{Grier2003} and references therein) has been used to create dynamic optical dipole potentials to manipulate BECs \cite{Boyer2006} using careful programming of the driving computer to work around the limitations of the SLM for BEC manipulation.  Other potential technologies for creating complex optical dipole potentials, although not yet demonstrated with condensates, include hybrid acousto-optic modulator/SLM systems \cite{Fatemi} and moving mirrors \cite{Beugnon2007}.

Unfortunately, all of these experimental technologies have failed so far to accomplish the goal of simultaneous dynamic manipulation and real time synthesis of arbitrary potentials while also maintaining conservative trapping that exhibits minimal heating and preserves phase coherence when necessary.  Here we present a solution to this challenge, with the experimental demonstration of a combination static plus time-averaged optical dipole potential that provides horizontal trapping, evaporation, and the ability to create and hold Bose-Einstein condensates in arbitrary time-averaged potentials.

 Some of the rich possibilities for creating complex time-averaged light fields have recently been demonstrated with a view to eventually trapping and manipulating condensates \cite{Schnelle, Houston2008}.  It appears, though, that the experimental implementation of time-averaged potentials with BECs is non-trivial.  Heating has been previously noted as an obvious challenge \cite{Onofrio}.  Further, when the scanning tweezer beam is also asked to provide axial confinement, we have found problems in dynamic manipulation due to condensate excitations and with thermal phase fluctuations \cite{Dettmer2001}, both consequences of the relatively weak axial confinement.  In our system we have therefore used a light sheet to provide tight confinement perpendicular to the axis of the focused optical tweezer beam.  The light sheet inhibits excitations in that dimension which could cause heating and it also reduces the trapping anisotropy to suppress phase fluctuations.  Our experiment exploits recent developments in inexpensive high N.A. aspheric optics to focus the tweezer to form a very tight trap, and also in fast arbitrary waveform generators with deep memory to synthesize arbitrary and dynamic waveforms for the two AODs which steer and modulate the tweezer beam.

\section{Experiment and Results}
\begin{figure}
\includegraphics[width=6.5in]{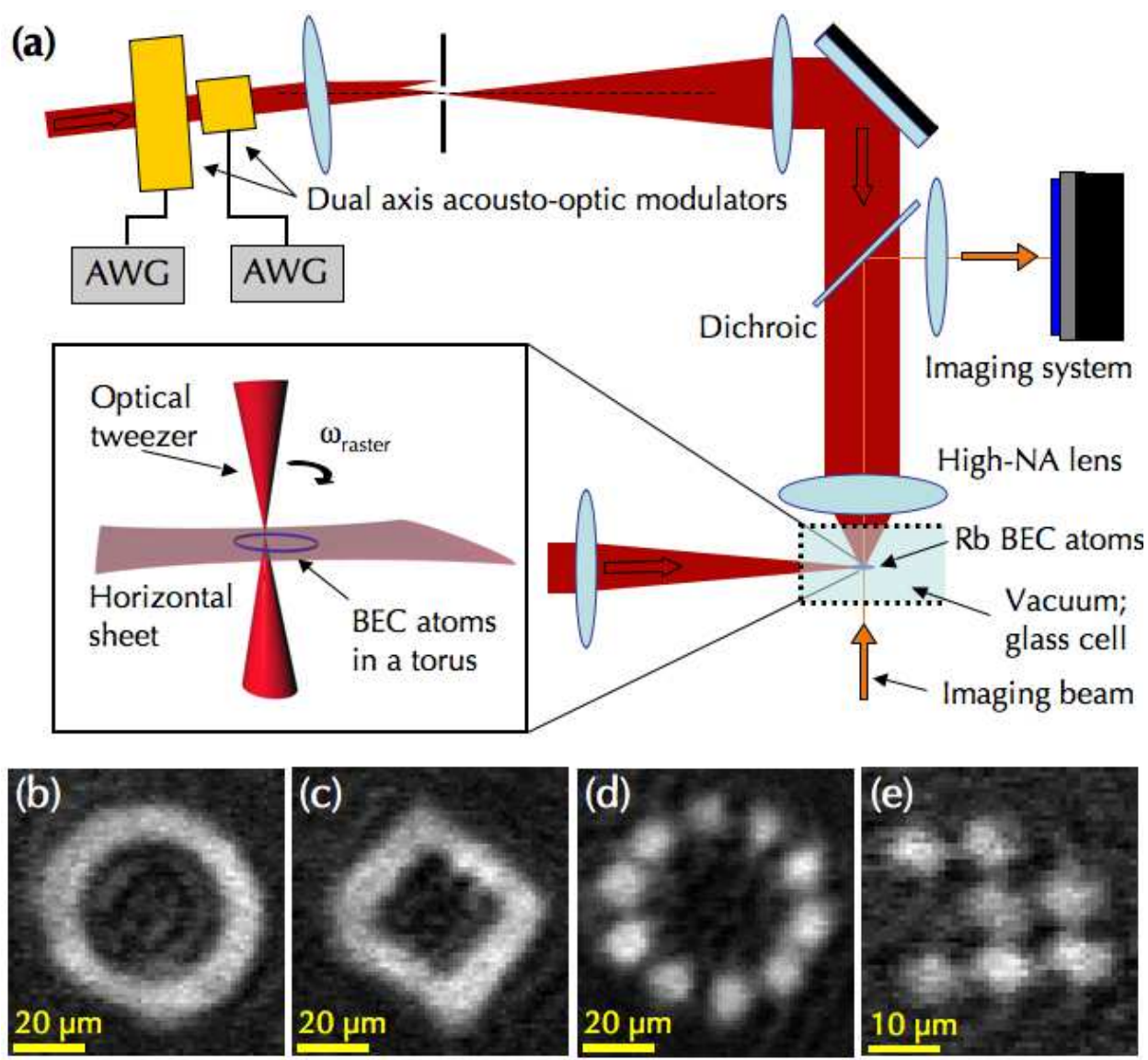}
\caption{\label{setup} (a) A single optical tweezer is focused by a 0.4 NA lens (Edmund Optics NT49-111) onto a sheet of light where atoms are evaporated and then condensed into time-averaged potentials.  Two acousto-optic deflectors (AOD) (IntraAction DTD-274HA6) are driven by two arbitrary waveform generators (AWG) (National Instruments PXI-5422) which control the location of the optical tweezer.  The inset shows an enlarged view of the trapping region. (b)-(e) are four examples showing in-situ absorption images of Bose-Einstein condensates formed in the crossed dipole trap with the optical tweezer painting a torus, diamond, ring of ten spots, and a three by three lattice with defects, respectively.}
\end{figure}

\subsection{Apparatus}
Our apparatus for producing BECs and cold atomic clouds has been described elsewhere \cite{Arnold2002}.  Figure\,\ref{setup} shows the elements added to this system to create and manipulate BECs in painted optical dipole potentials.  Briefly, $^{87}$Rb atoms in the $5^2S_{1/2}$ $|F = 2, m_F = 2\rangle$ ground state are first collected from a magneto-optical trap into a baseball configuration magnetic trap where they are evaporated to an average temperature of $400$\,nK after $30$\,s.  Typically $5 \times 10^5$ atoms are then transferred to a horizontal sheet of light ($\lambda = 1064$\,nm) with nearly $100\,\%$ efficiency.  The sheet forms an attractive potential on top of the magnetic trap as it is ramped on in $100$\,ms to a well depth of approximately $10\,\mu$K, with vertical and horizontal waists ($1/e^2$) of $7.4\,\mu$m and $300\,\mu$m, respectively.  The magnetic trap is then turned off quickly ($< 330\,\mu$s) and the power in the horizontal sheet is lowered in an exponential ramp to cool the cloud through evaporation.  A small magnetic field ($\sim 0.5\,$G) is applied throughout the experiment to maintain the atomic polarization.  All of the condensate images presented here are taken in absorption at detunings of 1 to 2 linewidths.  The imaging system has a resolution (FWHM) of $\sim5\,\mu$m, partly because the $\lambda=780\,$nm imaging beam passes through some of the tweezer beam optics which are optimized for $\lambda=1064\,$nm.

If the optical tweezer beam is kept switched off, Bose-Einstein condensates containing up to $1 \times 10^5\,$ atoms can be produced by reducing the depth of the sheet trap to a final value of $1\,\mu$K over a duration of $1000\,$ms.  The condensate cloud formed in the horizontal sheet has an aspect ratio of $1:30:40$, in the vertical and horizontal directions, respectively.  Lifetimes in this trap exceed $2\,$s and are limited primarily by background collisions and three-body recombination losses.  The chemical potential for condensates formed in the sheet trap is similar to the trapping energy in the tight dimension, and so the system is quasi-2D \cite{Gorlitz}, with ballistic expansion images showing a gaussian momentum profile along the tight axis.  The BEC in the sheet trap alone provides a $60\,\mu$m diameter canvas on which to paint a potential with the scanning tweezer beam.

To create condensates in complex optical dipole potentials, the time-averaged tweezer beam ($\lambda = 1064$\,nm, waist of $\sim2.5\,\mu$m,  and directed vertically downward through the sheet) is ramped on in $100\,$ms after only $400\,$ms of evaporation in the horizontal sheet.  The horizontal sheet power continues to drop for an additional $600\,$ms while the tweezer beam power is left constant.  In this way, atoms are evaporated in the combined dipole trap and allowed to condense directly into the potential created by the time-averaged tweezer.

\subsection{Static potentials}
Figure\,\ref{setup} (b)-(e) are in-situ absorption images of condensates formed in the painted potential.  We note that none of these relatively simple static geometries have previously been realized.  Toroidal BEC's have been made in magnetic traps with optical dipole barriers \cite{Ketterle1998, Ryu}, but in those cases the ratio of condensate thickness to central hole radius is large, as opposed to the thin condensate ring of figure\,\ref{setup}(b) (a note added at the end of \cite{Heathcote2008} reports loading of a BEC into a hybrid RF-dressed-magnetic/optical ring trap, but it gives no details or dimensions.)  BEC's have also been released into circular waveguides forming large ring traps \cite{Gupta2005, Arnold2006}, although toridal condensates have not yet been produced in these systems.  The $\sim5\,\mu$m condensate thickness seen in figure\,\ref{setup}(b) reflects our finite imaging resolution and the actual dimension is much smaller.  Wavefunctions computed numerically for our conditions have a radial thickness of less than $1\,\mu$m.

The number of condensate atoms depends on the volume generated by the potential geometry, but it is always less than the number of atoms condensed in only the horizontal sheet due to the smaller trapping volume.  As an example, the toroid in in figure\,\ref{setup}(b) contains around 5,000 atoms.  The power used in the tweezer depends on the geometry and complexity of the potential created by the scanning beam.  In the case of a simple $50\,\mu$m diameter toroid as shown in figure\,\ref{setup}(b), the tweezer power is approximately $1.5\,$mW, providing a time-averaged well depth of $390\,$nK for the toroidal potential and effective trapping frequencies in the radial and vertical directions of $\sim500\,$Hz.  The scanning tweezer beam paints the torus at a frequency of $4\,$kHz.

The lattice with missing sites in figure\,\ref{setup}(d) was created to show that this technique can easily make potentials not possessing any symmetry.  It should be possible to create lattice potentials such as these with many more sites.  We expect that the limitation will be the finite rise-time of the AOD, which for our device and a $4\,$kHz scanning frequency would permit 50 x 50 arrays.

While static versions of figure\,\ref{setup} (b) and (c) could easily be generated with discrete signal generators, and even (d), with the addition of modulation, the production of potentials with less symmetry (e.g. the lattice with defects (e)) and/or complex time evolution  (e.g. figure\,\ref{dynamics}) benefits from having a more flexible scanning system.  We achieve this by driving the AODs with fast (up to 100MS/s) arbitrary waveform generators with deep memory ($512\,$MB).  A computer calculates the RF waveform (centred on $24\,$MHz) corresponding to the desired time-averaged potential.  The AOD deflects and Doppler shifts the tweezer beam by an amount proportional to the RF frequency with an intensity proportional to the RF power.  We have chosen to follow a vector graphics paradigm because it has higher duty cycle than simple rastering.  The potential depth can be changed by velocity modulation, intensity modulation, or a combination of the two.  In the case of single continuous curves, we simply trace them out at constant scan velocity, reserving RF power changes to erect barriers, enhance flatness, etc., in the time-averaged potential. For multiple disconnected wells, we switch rapidly between them such that (ignoring modulation sidebands) only one frequency is present in the AOD drive at a time.  This avoids complications due to AOD efficiency changing when driven with multiple frequencies and artefacts when spots overlap due to beating between tweezers having slightly different frequencies.  It also avoids the symmetry constraints imposed by the fact that dual AODs convolve the two perpendicular deflections.  Although our system allows for potential normalization by changing the intensity of the beam as it scans, we have not found it necessary to invoke that option here.  Instead we use a diagnostic imaging system to monitor the scanning beam profile and optimize the deflection angle from each AOD to give the flattest response over the entire potential.  This typically results in small adjustments to the centre frequency of each channel, but is easily compensated with mirror mounts that redirect the beam to the location of the atoms at the centre of the horizontal sheet of light.  Limitations of the scanning frequency, $f_s$, are also analyzed using the diagnostic images.  For our setup, very high frequency scan rates, i.e, $f_s > 25\,$kHz, are not useful because they distort the potential and ultimately form modulation sidebands on the centre frequency of each channel.  The high frequency limitation is based on the response time of the AOD, which for our system is about $1\,\mu$s.  The radial trapping frequency of the time-averaged potential determines the limit for the lowest scan rate, which is typically on the order of $1\,$kHz.

\begin{figure}
\includegraphics[width=6.5in]{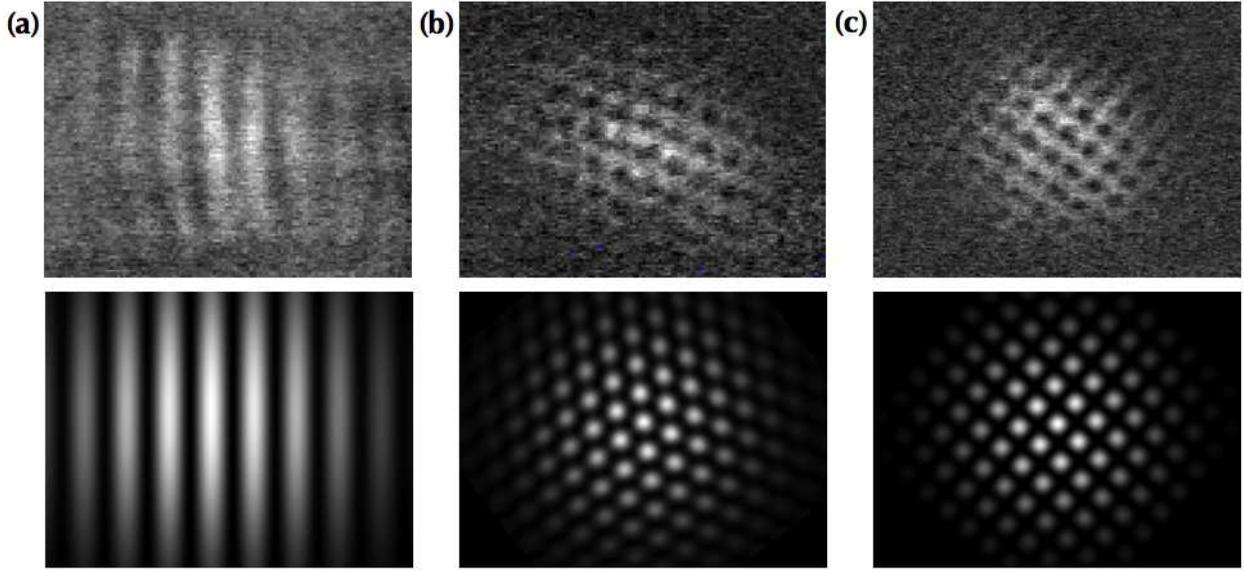}
\caption{\label{interference}  Time of flight absorption images taken $19\,$ms after atoms were released from the time-averaged potentials, and their simulations.  The scanning frequency is $4\,$kHz in all three cases. (a) Interference patterns from two spots separated by $5.4\,\mu$m. Fringe spacing is $16\,\mu$m. (b)  Interference pattern from three spots separated by $14\,\mu$m such that they create a equilateral triangle.  Fringe spacing is $6.7\,\mu$m. (c) Interference pattern from four spots separated by $11.4\,\mu$m such that they create a square. Fringe spacing is $7.9\,\mu$m.}
\end{figure}

It is of course important to verify that heating in the time-averaged potential is not significant, since that would result in thermal clouds rather than BEC's.   We have demonstrated that the objects produced by evaporation into the time-averaged potential are condensates by observing the interference fringes produced after ballistic expansion when the trap is turned off.  Some simple examples are shown in figure\,\ref{interference}.  We also observe no decrease in the number of condensate atoms after $2.5\,$s of optical trapping (the maximum duration possible with our current implementation), both in static potentials and in the dynamic potentials discussed in the following section.  The condensate lifetime in the time-averaged optical dipole potential is therefore presumably much longer than this time.

\subsection{Dynamic potentials}
\begin{figure}
\includegraphics[width=6.5in]{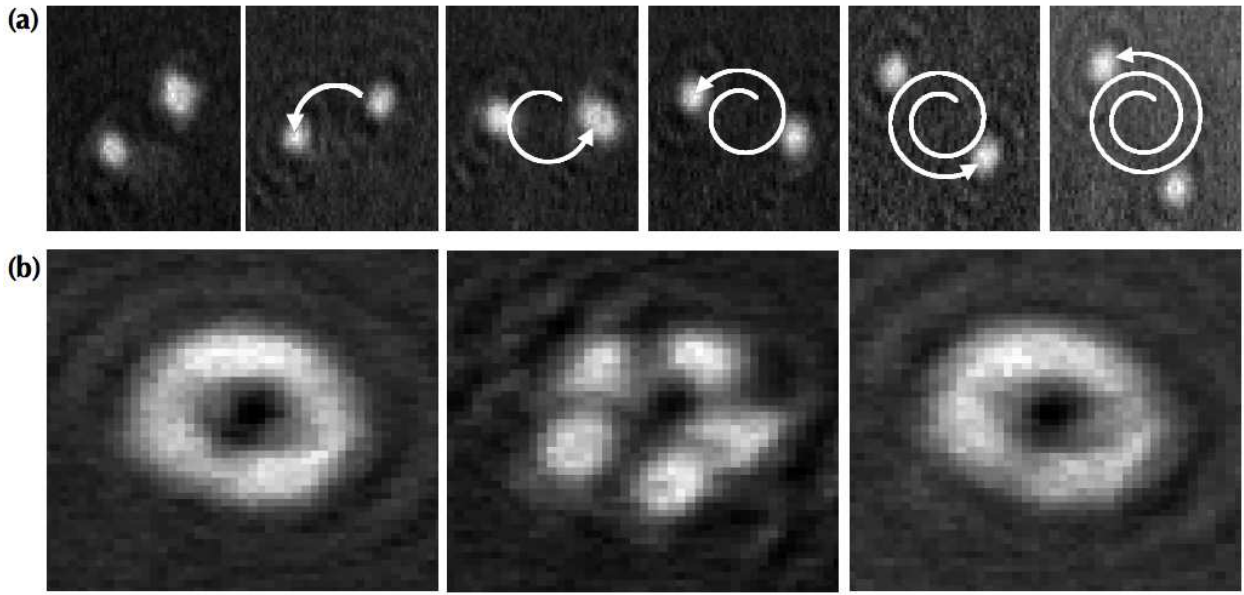}
\caption{\label{dynamics}  Two examples of dynamic manipulation of condensates. (a) A sequence of in-situ absorption images of two BECs initially separated by $10\,\mu$m and rotating at $15\,$Hz with increasing separation up to $25\,\mu$m. (b) A sequence of in-situ absorption images showing the transformation of a BEC trapped in a $20\,\mu$m diameter toroidal potential which is adiabatically converted into $5$ disconnected spots and then back into toroidal form.}
\end{figure}

Finally, in addition to producing arbitrary static potentials, a major advantage of the painted potential technique is that it can also drive complex dynamic changes in both the strength and the topology of the potential.  Figure\,\ref{dynamics} shows two examples.  The first demonstrates that multiple condensates can be moved in complex trajectories without heating.  The second (figure\,\ref{dynamics}(b)) is chosen to exhibit the control needed to realize the proposal of \cite{Dziarmaga2008} on quantum phase transition dynamics.  Here the scanning beam jumps at $4\,$kHz between toroidal and ring lattice potentials being scanned at $8\,$kHz, with time-varying amplitudes which implement the adiabatic transformation from to toroid to ring lattice and back to toroid.  The duration of each of the two adiabatic transformations is $200\,$ms.

One can easily imagine many other applications of these new possibilities for dynamic manipulation of BECs, such as rotating a deformed toroid to create persistent currents, or painting dynamically reconfigurable matter waveguide circuits.

\section{Conclusion}
In summary, we have shown that a rapidly-moving laser beam that paints a time-averaged optical dipole potential on top of a static light sheet is a straightforward method for creating BECs in arbitrary geometries, including toroids and ring lattices, and for manipulating them through complex dynamic changes in the potential.  These results are important because they establish for the first time that time-averaged optical dipole potentials can be a reliable and versatile tool for controlled experiments with Bose-Einstein condensates.  Finally, we note also that the technique is general and it should work with any sufficiently polarizable low-energy particles, including degenerate Fermi gases, conventional cold atoms, and even biological systems.

\ack{This work was supported by the U.S. Department of Energy through the LANL/LDRD Program.  We acknowledge inspiring conversations with Wojciech Zurek, Eddy Timmermans, and Aidan Arnold.}

\end{document}